\documentclass{aa}
\usepackage[varg]{txfonts}
\usepackage{siunitx}
\usepackage{graphicx}
\usepackage{hyperref}

\bibpunct{(}{)}{;}{a}{}{,}

\hypersetup{ 
    colorlinks,
    linkcolor=blue,
    citecolor=blue
}

\newcommand{\feI}{\ion{Fe}{i}~}
\newcommand{\caII}{\ion{Ca}{ii}~}

\newcommand{\theline}{\ion{Ca}{ii}~8542~\AA}
\DeclareRobustCommand{\ltau}{\ensuremath{\log \tau_{500}}}

\DeclareSIUnit\angstrom{\text {Å}}

\begin{document}

\title{Bifrost spectral inversions}
\subtitle{Fast non-LTE solar chromospheric diagnostics from 3D simulations}

\author{Elias R. Udn{\ae}s\inst{1,2},
        Tiago M. D. Pereira\inst{1,2},
        Ignasi J. Soler Poquet\inst{1,2},
        \and Luc Rouppe van der Voort\inst{1,2}
        }
\institute{Rosseland Centre for Solar Physics, University of Oslo, P.O. Box 1029 Blindern, NO--0315 Oslo, Norway
\and
Institute of Theoretical Astrophysics, University of Oslo, P.O. Box 1029 Blindern, NO--0315 Oslo, Norway}

\authorrunning{Elias R. Udn{\ae}s et al.}

\date{}

\abstract 
{Modern solar observatories produce vast amounts of detailed spectral data. To infer atmospheric parameters from these data, in particular for the solar chromosphere, requires immense computational resources. This problem is becoming more acute as spatial resolution improves, and faster methods are much needed to make sense of the data. Archive-based inversions provide a cost-effective way to infer atmospheric parameters from solar spectra; they involve building databases of synthetic spectra and atmospheric models. We extended this idea to a $k$-nearest neighbour inversion with non-LTE chromospheric spectra from a 3D radiative-magnetohydrodynamic simulation. Our database consists of 150 million \theline\ line profiles from a magnetically quiet simulation. We validated our method on a different simulation and then applied the database inversion to high-cadence observations taken with the Swedish 1-m Solar Telescope. We find excellent agreement between the observed and fitted spectra for almost the entire field of view. We recovered the chromospheric temperature and line-of-sight velocity and calculated the uncertainty in their inversions. The gas temperature has small uncertainties over a broad range of optical depths, and we recover a physically consistent solution, even beyond the line sensitivity region. Compared with the traditional inversion method, our approach is four orders of magnitude faster. Limitations and future improvements of the method are discussed.}
\keywords{Sun: chromosphere -- Methods: numerical -- line: formation -- radiative transfer}

\maketitle
\nolinenumbers

\section{Introduction}
Inferring chromospheric thermodynamics from spectropolarimetric observations is traditionally done via inversions \citep[see][and references therein]{de-la-Cruz-Rodriguez:2017aa}. However, both the higher volume and higher quality of the data that are acquired with modern observatories place immense demands on reliable and faster inversion routines. 

The main factor limiting high-throughput inversions of large field of view solar observations is the inherent non-locality of radiation in the chromosphere, which makes the inversion a global problem that has to be solved implicitly. Since calculations of atmospheric parameters are convoluted by non-LTE effects, inversions are computationally very expensive. As of today, it is too expensive to perform non-LTE inversions with full 3D radiative transfer on extended datasets \citep{Asensio-Ramos:2019aa}. The current norm for recovering 3D chromospheric information from observations is to perform non-LTE inversions independently for each pixel (so-called 1.5D). 

Much work has been done to alleviate the computational cost of inversions. Since the inversion problem is many-dimensional and global, it is a typical case in which machine learning algorithms work well. Some recent techniques where machine learning methods have sped up the inversions by several orders of magnitude include convolutional neural networks \citep{Asensio-Ramos:2019aa,Milic:2020aa}, neural fields \citep{Diaz-Baso:2025aa}, neural translation \citep{Asensio-Ramos:2025aa}, and Bayesian inference \citep{Diaz-Baso:2022aa}.

Other strategies for speeding up inversions rely on databases of synthetic spectra. Such databases are built from a set of model atmospheres and their spectra, and an inversion is taken as the archived model whose spectrum best reproduces a given observation. Several earlier works have used spectral archives for chromospheric inversion. Inversions based on grid searches have been performed for different spectral lines, including H$\alpha$ \citep{Molowny-Horas:1999aa,Schmieder:2003aa,Berlicki:2005aa}, \caII 8542~\AA\ \citep{Tziotziou:2001aa,Beck:2015aa,Beck:2019aa}, and \caII H \citep{Beck:2013aa}. Spectral archives have also been reduced with principal component analysis decomposition in order to further accelerate inversions \citep{Rees:2000aa,Lopez-Ariste:2002aa,Casini:2003aa}. More recently, \citet{Sainz-Dalda:2019aa,Sainz-Dalda:2026aa} used representative profiles from spectra inverted with STiC \citep{de-la-Cruz-Rodriguez:2019aa} to speed up analyses of large datasets from the Interface Region Imaging Spectrograph \citep[IRIS;][]{De-Pontieu:2014aa}. These representative profiles act as a lookup table when mapping observations to thermodynamic variables. With spectra synthesised from a 3D radiative-magnetohydrodynamic (rMHD) simulation, \citet{Riethmuller:2017aa} used database inversions of photospheric spectra as an initial condition for a simulation in order to obtain a physically consistent magnetohydrodynamic model that is similar to an observation. However, no studies have used 3D rMHD simulations with chromospheric conditions to infer an atmosphere's thermodynamic state.

\begin{figure}
    \resizebox{\hsize}{!}{\includegraphics{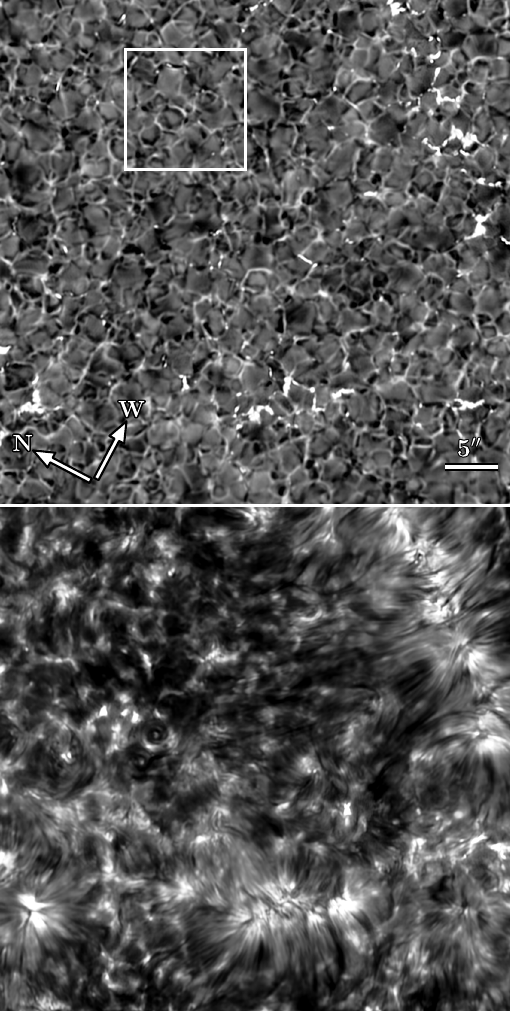}}
    \caption{A moment of particularly good seeing in the observed region. \textit{Top}: Red wing at +\qty{1.2}{\angstrom} from the \theline\ line centre. The white rectangle shows a quiet patch of the observation and is the region we later use to compare our inversion with the STiC inversion. \textit{Bottom}: \theline\ line centre. The images were taken at 2014-06-24T09:15 and are cropped to $48\farcs7 \times 48\arcsec$.}
    \label{fig:observation_overview}
\end{figure}

In recent years, 3D rMHD simulations have attained a high degree of realism \citep{Pereira:2013ab} and reproduce the solar spectrum to a high confidence \citep[e.g.][]{Witzke:2024aa,Ondratschek:2026aa}. They can also include more realistic physics than the 1D model atmospheres traditionally used in spectral inversions \citep[e.g. reproducing line widths without using microturbulence;][]{Asplund:2000aa}. In this study we employed a 3D rMHD simulation to infer chromospheric quantities in a solar observation by comparing observed spectra pixel by pixel with spectra synthesised from a Bifrost simulation \citep{Gudiksen:2011vu}. We validated our method by inverting a snapshot from the publicly available enhanced-network simulation \citep{Carlsson:2016aa}, and demonstrate our database inversion on a CRISP observation \citep{Scharmer:2008aa}. By associating an observation with not one but several synthetic spectra, we also estimated both the uncertainty and the degeneracy of the atmospheric parameters in the inversion. We extended this to an uncertainty analysis of each atmospheric parameter as a function of optical depth. Lastly, we compared our results with a traditional non-LTE inversion using STiC.

\section{Data}

\begin{figure}
    \resizebox{\hsize}{!}{\includegraphics{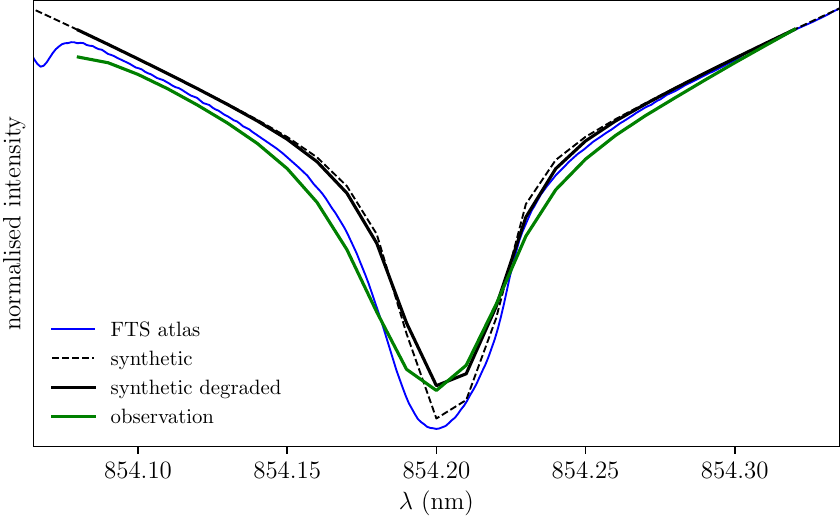}}
    \caption{Average \theline\ spectrum from the observations and the simulation. The blue line is taken from the FTS atlas. The black lines represent the synthetic spectra: the dashed line is averaged over the non-degraded spectra, and the solid line is averaged over the spectra degraded to the SST spatial and spectral resolution. The green line is the mean spectrum from the SST observation. All spectra are normalised by the local continuum at $\lambda_0 + \qty{1.2}{\angstrom}$, which corresponds to the last wavelength point of the SST observation.}
    \label{fig:avg_spectrum}
\end{figure}

\begin{figure}
    \resizebox{\hsize}{!}{\includegraphics{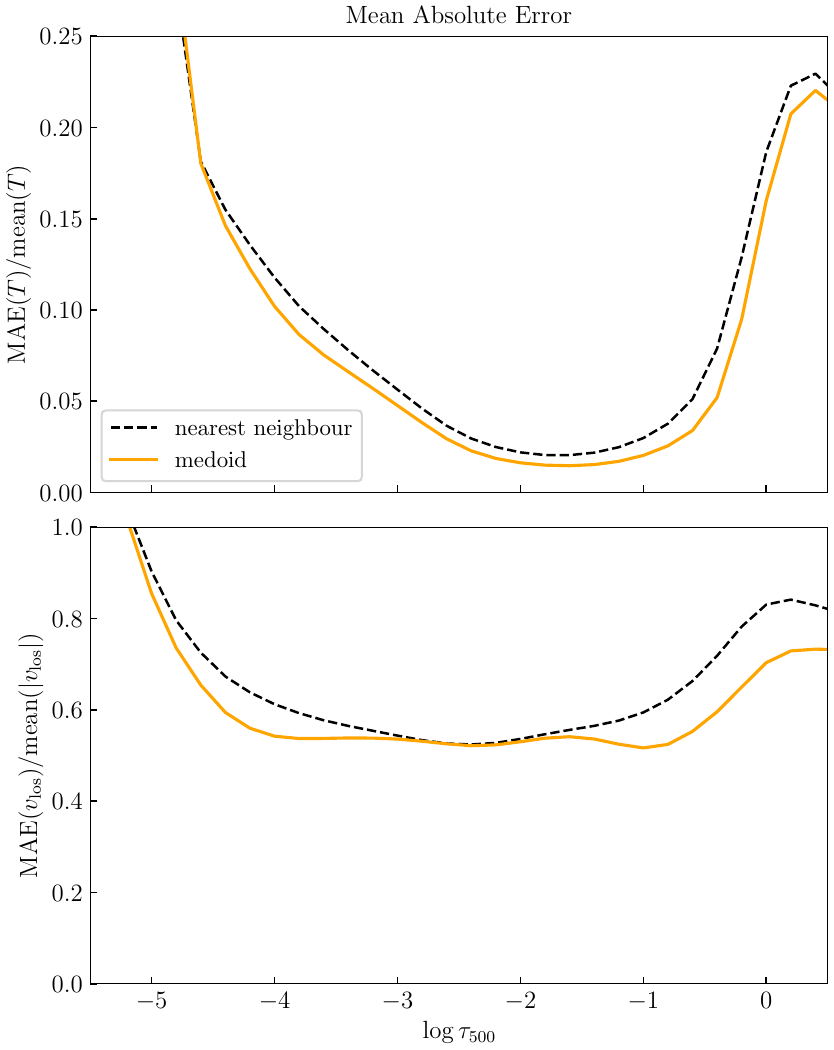}}
    \caption{MAE for the BiSPEC inversion of the temperature ($T$) and line-of-sight velocity applied to the enhanced network simulation for nearest neighbour interpolation and the $k$-NN medoid. The MAE is normalised with the mean of each quantity per each optical depth.}
    \label{fig:MAE}
\end{figure}

\begin{figure*}
    \includegraphics[width=17cm]{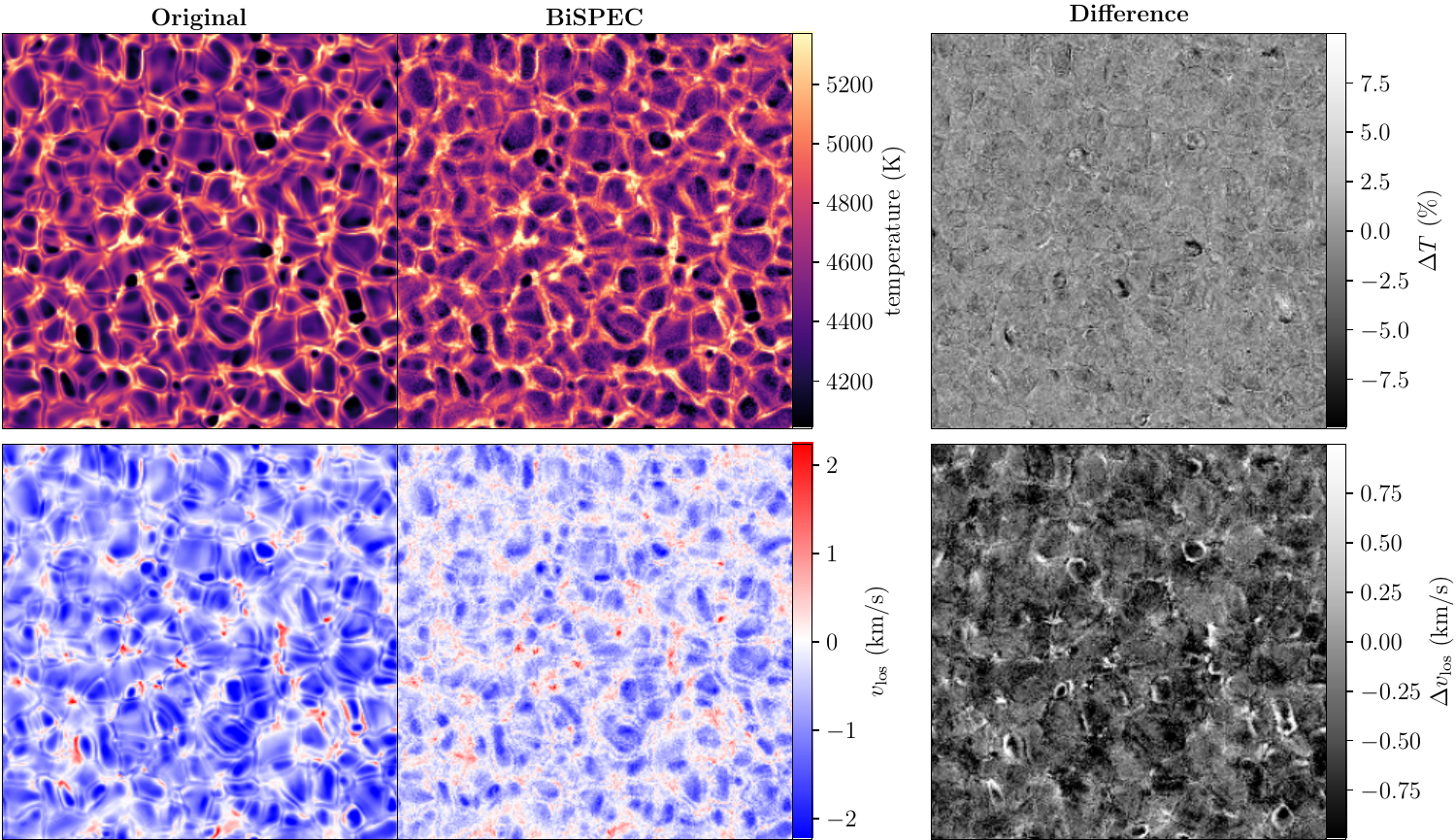}
    \caption{Comparison between the enhanced network simulation (original) and the BiSPEC inversion at $\ltau = -1.8$. The difference is taken as the original value minus the value of the inversion, and the relative difference for temperature is normalised by the ground truth value. In the line-of-sight velocity, negative (positive) values correspond to upflows (downflows) in the atmosphere.}
    \label{fig:sim_comp}
\end{figure*}

\begin{figure*}
    \sidecaption
    \includegraphics[width=12cm]{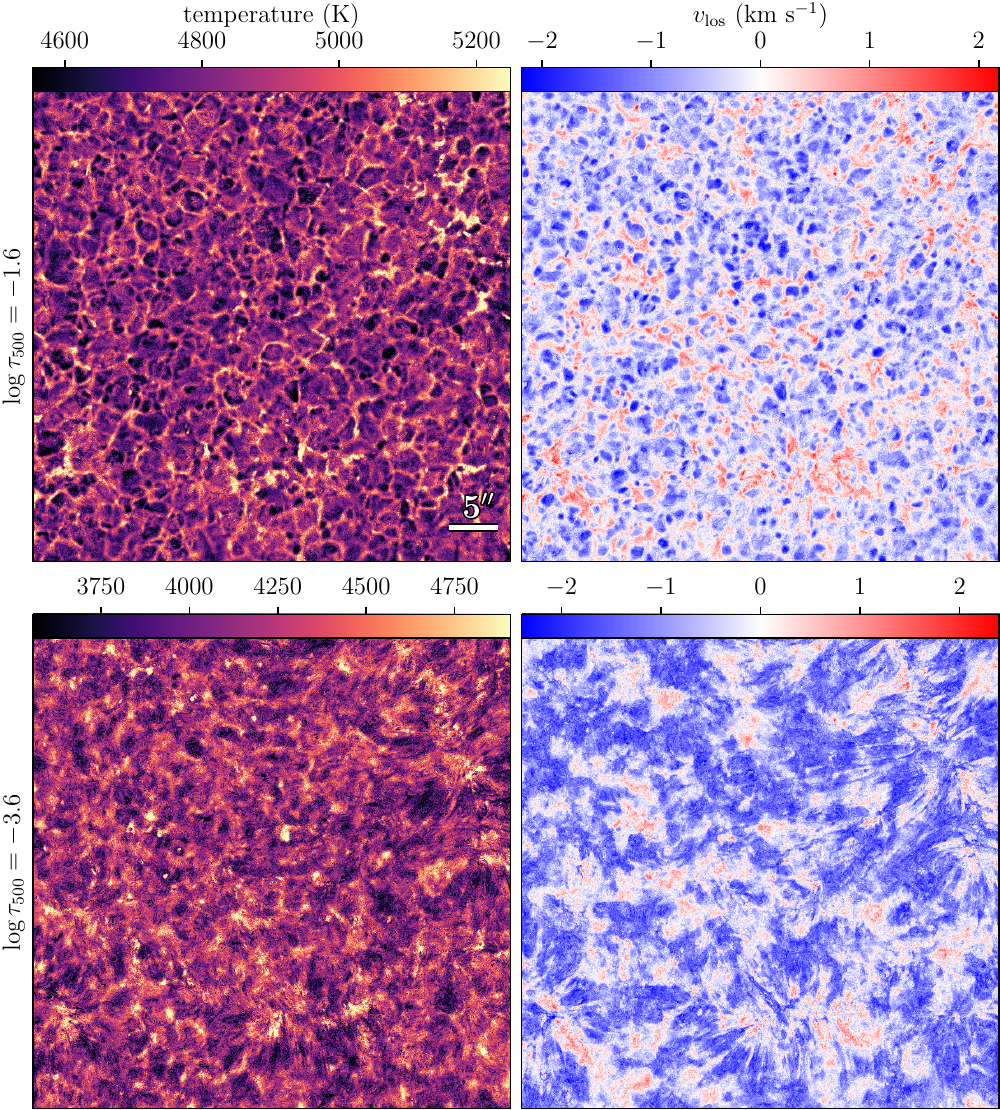}
    \caption{Inversion of the \theline\ observation at 2014-06-24T09:15 shown in Fig.~\ref{fig:observation_overview}. The columns show the inferred temperature and line-of-sight velocity ($v_\mathrm{los}$) at $\tau_{500}$ isosurfaces. \textit{Top row}: $\ltau = -1.6$. \textit{Bottom row}: $\ltau = -3.6$. Positive velocities (red) correspond to downflows, and negative velocities (blue) correspond to upflows.}
    \label{fig:inversion_tau500}
\end{figure*}

\subsection{Observations}

We studied observations of the \theline\ spectral line from a small trans-equatorial coronal hole region, using the same dataset as \citet{Moe:2024aa}. Data were acquired with the CRISP instrument \citep{Scharmer:2008aa} at the Swedish 1-m Solar Telescope \citep[SST;][]{Scharmer:2003aa} on 24 June 2014. We processed the data using the CRISPRED data reduction pipeline \citep{de-la-Cruz-Rodriguez:2015ab}, which uses the image restoration technique multi-object multi-frame blind deconvolution \citep[][]{Van-Noort:2005aa}. The data were acquired between 08:27 UT and 09:42 UT, giving a time series of 1 hour and 15 minutes. Seeing conditions were excellent throughout the dataset, and with adaptive optics corrections \citep{Scharmer:2024aa} and after multi-object multi-frame blind deconvolution image restoration the spatial resolution of the data was close to the diffraction limit of the telescope at $0\farcs 18$ for 8542~\AA.

The field of view of our observation was close to disk centre, centred at heliocentric Cartesian coordinates $(x,y) = (-119\arcsec, -106\arcsec)$; the viewing angle was $\mu \approx 0.99$. The dataset consists of H$\alpha$ and \theline\ Stokes I spectra, and one line-position with full Stokes in the wing of \feI 6302~\AA. Here, we used only the \theline\ data, which were sampled with 25 evenly spaced line positions around the line centre with 0.1~\AA~ separation. Due to the sparse sampling of the other lines in the programme, the temporal cadence of the data is high at \qty{11.5}{s}. In Fig.~\ref{fig:observation_overview} we show an overview of our observation in two different wavelengths taken at a moment when the seeing conditions were particularly good: Fried's parameter ($r_0$) measured for the whole atmosphere peaked at \qty{20}{cm} ($r_0 > \qty{40}{cm}$ for the atmosphere near the telescope; \citealt{Scharmer:2019aa}). In the line wing there is a bright network region extending across the field of view; it is seen as chromospheric spicule bushes in the line core. 

\subsection{Simulation}
The simulation we used in this work is a quiet-Sun patch produced with the Bifrost code \citep{Gudiksen:2011vu}. Our simulation is named \texttt{ch024031\_by200bz005}, where \texttt{ch} stands for coronal hole, \texttt{024} comes from the \qtyproduct{24 x 24}{\mega\metre} horizontal extent, and \texttt{031} is the \qty{31.25}{km} grid spacing in $x$ and $y$. The simulation box is discretised into $768^3$ grid cells with a vertical domain spanning the convection zone from \qty{2.5}{Mm} beneath the surface, the photosphere, the chromosphere, and ending in the corona at \qty{14.3}{Mm} above the surface. In the $z$-direction, the grid spacing is not uniform; it is finer in the chromosphere than in the photosphere and the corona. The average vertical grid spacing is \qty{21.9}{\kilo\metre}.

In the simulation name, \texttt{by200bz005} indicates the magnetic configuration. The simulation was run with an initial horizontal field $B_{y,0}=\qty{20}{mT}=\qty{200}{G}$ but the field that entered the domain was much smaller than this value. At the beginning of the simulation, a field with a mean signed $B_z$ of \qty{0.5}{mT} was added; at the photosphere the simulation has a mean signed magnetic field that is close to \qty{4.5}{mT} during the entire time series.
With this mean signed magnetic field, the simulation is magnetically quiet. There are mixed polarities in the photosphere, but the coronal magnetic field is vertical \citep[as can be seen in Fig.~1 of][]{Finley:2022aa}. The average effective temperature of the simulation is \qty{5754}{K} through the 2 h and 45 min duration that we analysed, with a standard deviation of \qty{8}{K} over that time.

\begin{figure}
    \resizebox{\hsize}{!}{\includegraphics{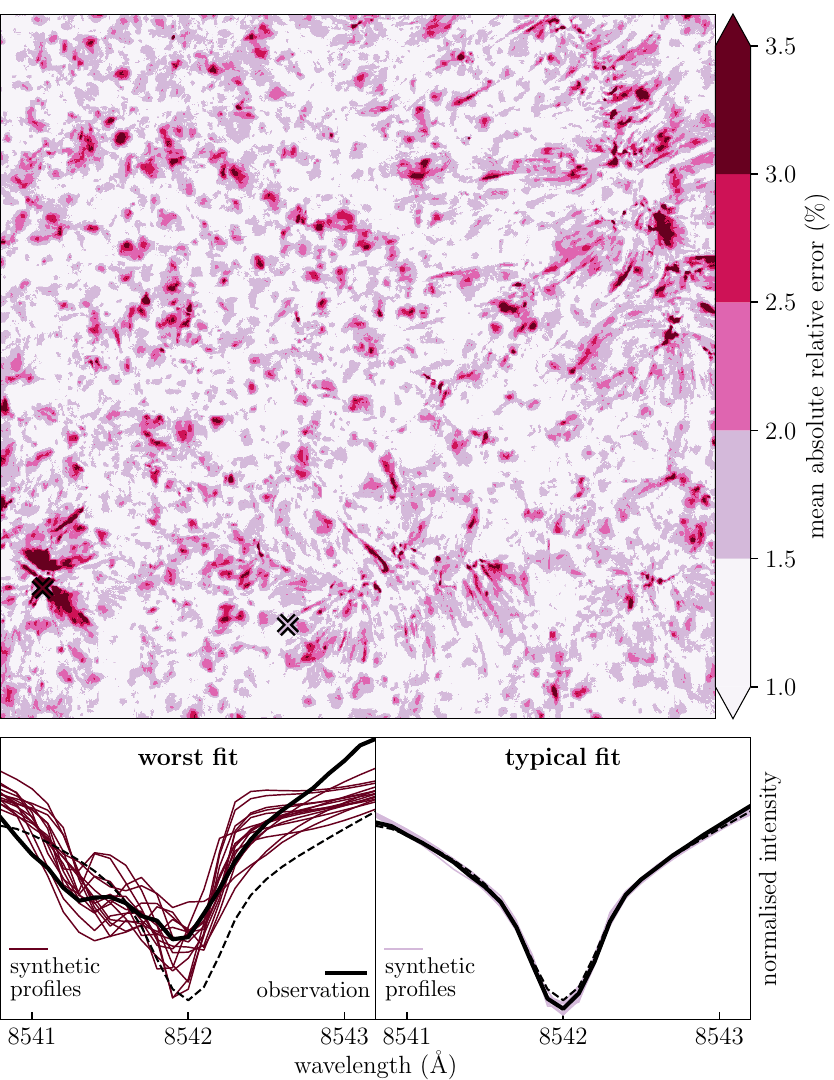}}
    \caption{Mean of the absolute relative error computed between the $k$ closest neighbours and the observation shown in Fig.~\ref{fig:observation_overview}. \textit{Top row}: Wavelength-averaged relative error of the fitted spectra. The crosses show the locations of the worst (red) and typical (pink) fitted pixels in the field of view. \textit{Bottom row}: Spectra with the worst and typical fits. The worst fit has a mean relative error of 9.4\%, and the typical fit lies at the median value of the mean relative error of 1.5\%. Solid black lines show the observed spectra in the marked pixels, and the coloured lines show the synthetic spectra used in the inversion of the pixels. The dashed black lines show the average spectrum from the observation.}
    \label{fig:inversion_err}
\end{figure}

\subsection{Synthetic spectra}
We synthesised \theline\ data from the full field of view for 999 snapshots of the simulation. The snapshots each have a cadence of \qty{10}{s}, for a total of 2 h and 45 min of solar time. The radiative transfer calculations were performed column by column with the non-LTE spectral synthesis code RH1.5D \citep{Uitenbroek:2001wf,Pereira:2015wv}. We modelled \caII as a five-level atom plus continuum, with the assumption of complete redistribution for all lines \citep[which is a good assumption to model the \caII infrared triplet lines,][]{Shine:1975aa,Uitenbroek:1989aa,Bjorgen:2018aa}. The spectral synthesis was performed without including Zeeman broadening. We performed a test by including Zeeman broadening on synthetic spectra from a single simulation snapshot, and found a negligible impact on the broadening of the lines.

To speed up calculations, we calculated spectra at every second pixel in $x$ and $y$ from the simulation. This is the same strategy employed in \citet{Udnaes:2026aa} and does not affect the effective spatial resolution of the data \citep[we refer to the discussion in][]{Moe:2022aa}. The spatial sampling of our spectra is then \qty{62.5}{\km}, or $0\farcs086$ as observed at a distance of \qty{1}{AU}.

After the \caII populations were converged, we re-calculated the emergent intensity of \theline\ with the \texttt{Muspel.jl} radiative transfer library \citep{muspel_v026}, along the vertical line of sight (viewing angle $\mu=1$, i.e. disk-centre observations). This step accounts for the effects from calcium isotopes in the solar atmosphere, which gives the inverse-C shape in the bisector of \theline\ \citep{Leenaarts:2014aa}. In our approach we assumed that the departure coefficients are the same for different isotopes, which \citet{Ondratschek:2026aa} show is a good approximation for the \theline\  profile shapes. We synthesised the line profiles from \qtyrange{-3}{+3}{\angstrom} around the line centre in \qty{0.1}{\angstrom} steps, which covers the spectral range of our observations.

We also used \theline\ spectra from one snapshot of the publicly available enhanced-network simulation \citep{Carlsson:2016aa}. These spectra were used only to validate our method, and are therefore not part of the synthetic spectra database. The validation data were synthesised with the same procedure as described above.

In order to compare the synthetic spectra with our observation, we convolved the spectra with the spatial and instrumental profile (spatial and wavelength) of CRISP. Since the SST spatial resolution at \qty{8542}{\angstrom} is close to the spatial resolution of our synthetic spectra, only a slight Gaussian blur ($\sigma = 0.9$ pix) was applied to the spectra  to account for the spatial degradation. The most significant degradation effects come from convolving the synthetic spectra with the transmission function of CRISP, whose full width at half maximum at 8542~\AA\ is about 0.1~\AA\ \citep{de-la-Cruz-Rodriguez:2015ab}. We compare the average synthetic spectrum (before and after convolution) with the SST observation and the Fourier Transform Spectrometer (FTS) atlas \citep{Wallace:1998} in Fig.~\ref{fig:avg_spectrum}. The normalised average synthetic spectrum has a narrower line core than the average SST observation and the FTS atlas. Getting a good match between the mean observed and synthetic line widths is a challenge for most 3D rMHD models. Earlier models predicted much narrower \theline\ profiles \citep{Leenaarts:2009aa}, but more recent simulations produce a better match \citep[e.g. \texttt{ch012023} in][]{Moe:2024aa}, and others exhibit mean profiles that are broader than the observations \citep{Ondratschek:2026aa}. While it is obviously desirable that the mean profiles match between observed and synthetic, our focus is on spatially resolved spectra, and it is clear that there are many locations in the simulations that closely match individual observed spectra \citep[as was already noted by][with several clusters, even from different simulations, having shapes similar to observations]{Moe:2024aa}.

\section{Inversion based on $k$-nearest neighbours}
Our inversion is based on a $k$-nearest neighbour ($k$-NN) procedure. A \theline\  observation is inverted by comparing it with the $k$ closest synthetic spectra in our database, where the distances between the observation and database spectra are calculated with the Euclidean norm. In the distance norm, all wavelength points are weighted equally. The nearest neighbours are then found by using a k-D tree to speed up distance calculations. 

If each neighbour reproduces an observation, every neighbour has an atmosphere that is consistent with the observed spectrum and can be taken as an inversion.  However, as we show later, we find that using the $k$-NNs to inform the inversion improves the solution. The inversion is taken as the medoid of the $k$-NNs, which represents the most typical atmospheric stratification in the set of neighbours. Our method is similar to $k$-NN regression except that the medoid is an actual simulation column, which ensures consistency between our inversion and observation.

In $k$-NN regression, weights can be assigned to the contributions of each neighbour, which are related to the distance from each neighbour. In our work we found that for values of $k$ of the order of ten, all neighbours used in the inversion had approximately equally small distances and were equally good representations of the observed spectra. Therefore, we did not use weights in our inversion, and we counted each neighbouring synthetic spectrum equally. For a more detailed description of the inversion procedure, see Appendix~\ref{app:k-nn}.

\begin{figure}
    \resizebox{\hsize}{!}{\includegraphics{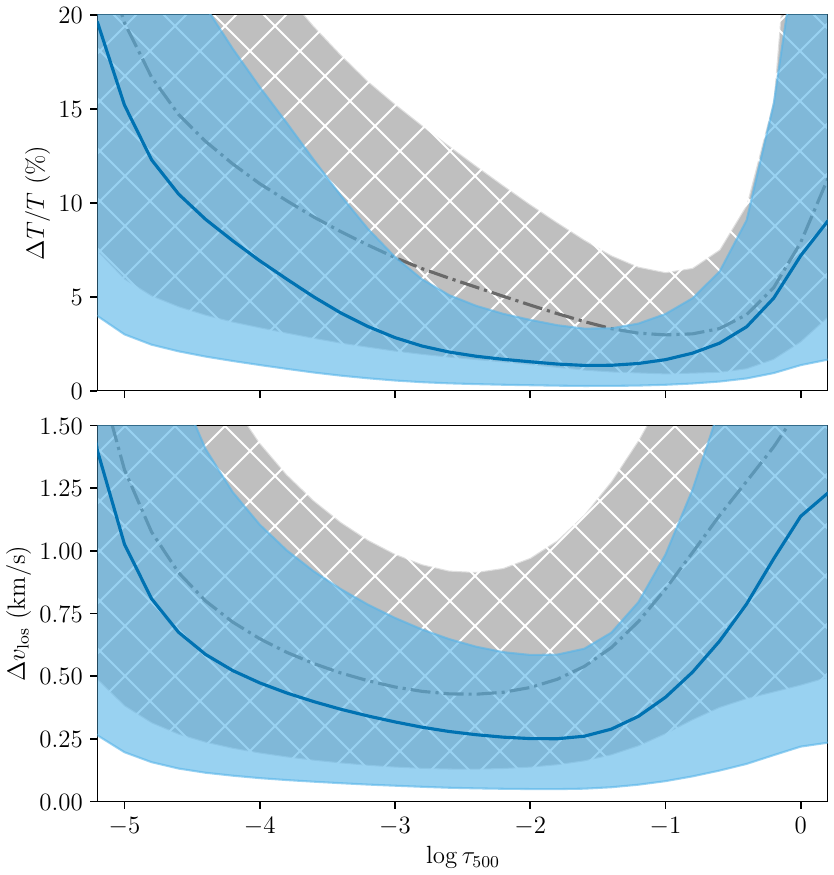}}
    \caption{Spread of the inverted quantities compared with the spread of the same quantities in the simulation. The solid blue lines show the median of the absolute deviation over all inverted pixels, and the shaded blue regions show the area between the 16th and 84th percentiles of the absolute deviation. The dash-dotted black lines show the median absolute deviation of each quantity over all simulation columns, and the hashed grey regions show the area between the 16th and 84th percentiles of the simulation's absolute deviation. Relative values were obtained by normalising by the median quantity at each optical depth. \textit{Top}: Relative uncertainty in temperature. \textit{Bottom}: Absolute uncertainty in the line-of-sight velocity.}
    \label{fig:inversion_deviation}
\end{figure}

The main advantage of using a $k$-NN procedure instead of nearest-neighbour interpolation, is that we obtain $k$ different solutions to the radiative transfer equation, which makes it possible to estimate uncertainties in the inversion. Via visual inspection of the data, we found that using $k=15$ generally gave good fits between an observed spectrum and all its neighbouring synthetic spectra, and gives us a small sample space that we used to estimate the distribution of atmospheric parameters. A short discussion on the differences between the $k$-NN inversion and nearest-neighbour interpolation is given in Appendix~\ref{app:k-nearest}. Throughout the rest of this paper we refer to our inversion method as BiSPEC, from Bifrost SPECtral inversion.

Our database comprises synthetic spectra from every fourth pixel in $x$ and $y$ from 999 simulation snapshots. This is a quarter of the spectra that we synthesised and approximately \qty{37}{million} of the almost \qty{150}{million} synthetic spectra we have available. Each database spectrum has an associated simulation column, which is interpolated onto a common $\tau_{500}$ optical-depth scale. Transforming the height scale requires interpolation, and in this case we performed the interpolation of atmospheric quantities with a cubic spline. 

\section{Results}

\subsection{Validating on the enhanced network simulation}
We validated our inversion on synthetic spectra calculated from a single snapshot of the enhanced network simulation. The spectra were compared in the wavelength range of $\pm1.2$ \AA\ around the line centre, without any degradation performed on the spectra. Figure~\ref{fig:compare_k} shows the line-centre intensity of this frame along with a nearest-neighbour reconstruction from the spectral database.

Figure~\ref{fig:MAE} shows the mean absolute error (MAE) between the original and the inverted gas temperature and line-of-sight velocity ($v_\mathrm{los}$) of the enhanced network simulation. The MAE is calculated per optical depth and normalised by the average quantity at each optical depth. We show the MAE for both the nearest-neighbour interpolation and the $k$-NN medoid (or BiSPEC) inversions, and observe that the medoid has a lower MAE throughout the domain. 

The errors in our temperature inversions are lowest at the depth $\ltau = -1.8$. We compare the inversions with the ground truth at this depth in Fig.~\ref{fig:sim_comp}, where we plot the inversion as the medoid of the $k$-NN. The figure shows that temperatures are well reconstructed, with small differences of the order of a few percent. The inverted line-of-sight velocity has larger differences compared with the original simulation data, but visual inspection of the maps of velocities reveals several similarities between the ground truth and BiSPEC inversion, such as the granulation pattern. In general, the inferred velocities share the same sign with the ground truth (i.e. the direction of the flow is conserved), but the inversion shows smaller amplitudes. The difference plot shows locations in the granules with a negative difference (under-estimated upflows) and bright regions in the inter-granules of around \qty{1}{km.s^{-1}} (under-estimated downflows). The small amplitudes in the inferred line-of-sight velocities are an artefact of the $k$-NN inversion, and nearest neighbour interpolation yields slightly stronger flows (shown in Fig.~\ref{fig:inversion_nearest_neighbours}).

\subsection{Inverting a single spectral scan}
We performed the inversion on the observation taken at 2014-06-24T09:15, corresponding to the frame shown in Fig.~\ref{fig:observation_overview}. This frame is slightly cropped from the original observation and consists of $854 \times 842 \approx 720,000$ pixels. The inversion of the full 2D spectral scan took approximately 15 minutes using 32 cores on an Intel Xeon 2.10~GHz machine. From the synthetic spectra database around 1.2 million spectra were used to invert this frame, with many synthetic spectra reused for different pixels. The reconstructed narrow-band images show virtually no difference from the observations. We present the atmospheric inversion of the observation in Fig.~\ref{fig:inversion_tau500}, which shows the temperature and line-of-sight velocity at two different optical depths, $\ltau = -1.6$ and $\ltau = -3.6$. These depths are locations where the \theline\ Stokes I response functions show sensitivity to temperature and line-of-sight velocity \citep[see e.g.][]{Quintero-Noda:2016aa}. 

At $\ltau = -1.6$ the atmospheric quantities show clear structures that correlate with the intensity at the \theline\ line wing. Temperature follows the reversed granulation pattern with lower temperatures above the granules, and higher temperatures above the intergranular lanes. The line-of-sight velocity also follows the convection patterns. The network region at this depth is clearly seen as hotter, less dense and with stronger downflows. However, we advise against drawing strong conclusions from our inversion in the network, since our simulation does not contain network-like features. Towards the chromosphere, at $\ltau = -3.6$, the atmospheric variables are more diffuse. This depth corresponds to the average depth of the temperature minimum of the simulation. The material here is mainly upflowing and relatively cool. 

\begin{figure}
    \resizebox{\hsize}{!}{\includegraphics{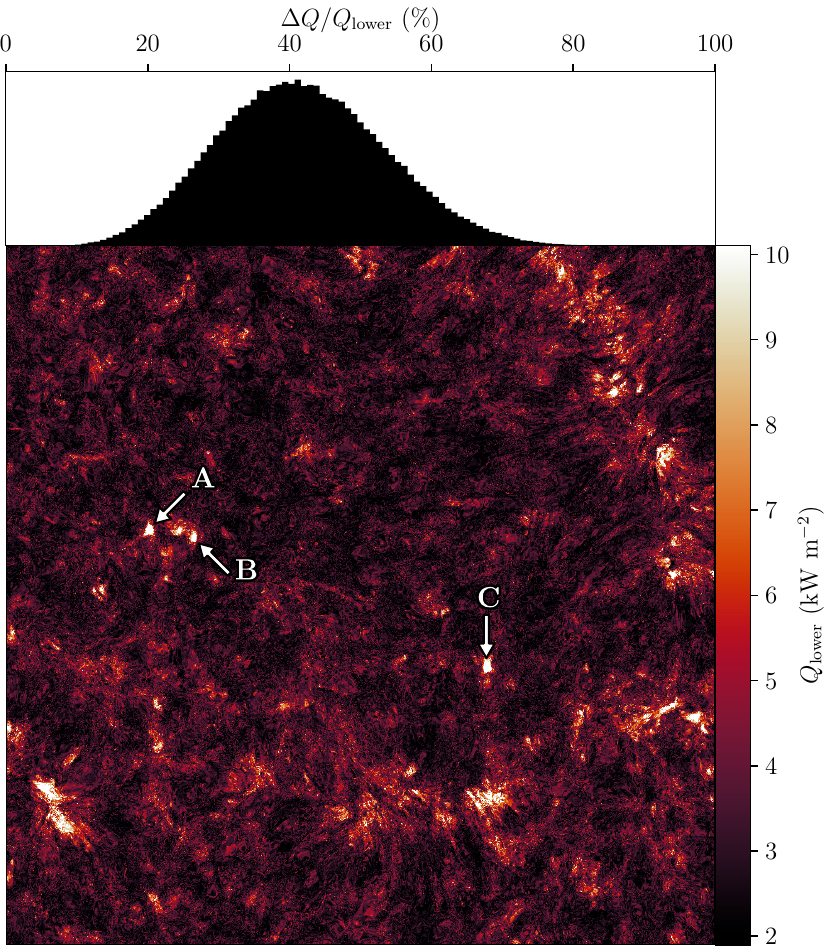}}
    \caption{Estimated chromospheric heating in the field of view shown in Fig.~\ref{fig:observation_overview}. The chromospheric heating is the sum of the dissipative viscous and Joule heating processes integrated over the column-mass range \qtyrange{1e-1}{1e-3}{kg.m^{-2}}. \textit{Top}: Histogram showing the distribution of uncertainties in our inferred heating. \textit{Bottom}: Inferred chromospheric heating in the field of view. The letters (A, B, and C) highlight three bright blue grains with elevated heating.}
    \label{fig:heating}
\end{figure}

Figure~\ref{fig:inversion_err} shows the relative error of the fitted profiles averaged over wavelength. There is generally a close match between each observation and its best-fitting synthetic profiles. The median of the mean relative error is 1.5\%, with fewer than one in one hundred pixels having an error larger than 3\%. Figure~\ref{fig:inversion_err} also shows a typical- and the worst-fitted observational spectra. The worst fit has a mean relative error of 9.4\%, and is located in a network patch. While 9.4\% may not seem to be a large error, the figure demonstrates that this is a poor fit. The synthetic profiles of the worst fit deviate significantly from the observation, which has another absorption component in the blue wing that none of our synthetic profiles have. 
By visual inspection of the network region, we saw that the blueshifted absorption component was caused by a fast moving feature that was visible for only two frames. This line profile could suffer from continuum-induced changes during the acquisition time of the Fabry-Perot scans \citep[see][]{Schlichenmaier:2023aa}. Other than such double-component line profiles, which are rare in the data, the network regions generally have profiles that the simulation does not reproduce well. In the network, the most common feature for poorly fitted profiles is elevated line wings. Some poorly adapted profiles are elevated in both the wings and line cores, such as the profile shown in Fig.~\ref{fig:inversion_err}. Except for the double-component absorption however, our synthetic profile database provides good qualitative matches for all observed profiles, even those with elevated wings. An interesting point to note is that the best-fitting profiles usually come from intergranular lanes while profiles at granule centres have larger errors.

\subsection{Estimating uncertainties}
We used the simulation columns of the 15 $k$-NN synthetic spectra to calculate the spread in the inversion of atmospheric parameters. The spread was calculated for an atmospheric quantity at each depth point in the $\tau_{500}$ grid as the absolute deviation from the median,
\begin{equation}
    \Delta X = \left | X - \tilde{X} \right |\,,
    \label{eq:absolute_deviation}
\end{equation}
where $X$ is the atmospheric quantity and $\tilde{X}$ is the median of the same quantity. This spread is then used as an approximation of the uncertainty in the inversion.

We calculated the spread over all pixels in the inversions of the frame presented in Fig.~\ref{fig:observation_overview} and show their distributions in Fig.~\ref{fig:inversion_deviation}. This figure shows the 16, 50 (median), and 84 percentiles of the absolute deviation of the temperature and line-of-sight velocity. It also compares the spreads of the inverted quantities with the spread of the same quantity in the simulation. The latter is important to use as a comparison since the uncertainty of our inversion is bound by the distributions of each quantity in the simulation. For example, if the temperature in the simulation at a fixed optical depth were constant, then the uncertainty in the temperatures in the inversion would also be zero at this optical depth.

For the uncertainties in temperature presented in Fig.~\ref{fig:inversion_deviation}, we normalised the absolute deviations with the median temperature to obtain a relative uncertainty. For the line-of-sight velocity, which oscillates around zero, we show the absolute uncertainty instead. The temperature and line-of-sight velocity have the smallest uncertainties in the lower atmosphere where $-3 < \ltau < -1$. The spread of temperatures in this region is low,  typically 2\%. The typical spread in the line-of-sight velocity is approximately \qty{0.3}{km/s} in the same region. 

The range of optical depths where the inversions are well estimated follows the response functions of \theline, which for the wings and core of \theline\ are sensitive roughly from depths spanning from $\ltau = -1$ to $\ltau = -5$ \citep{Quintero-Noda:2016aa}. However, uncertainties also scale with the overall spread of each quantity in the simulation at a given depth. The simulation has, in general, narrow ranges of velocities and temperatures around $\ltau = -2$: it varies less in temperature and velocity at these depths. Still, the 84th percentile for the inverted temperature is below the simulation's median spread between $-3 < \ltau < -1.5$, which signals that this quantity is well constrained by the \theline\ data at these depths. The spread in the inferred line-of-sight velocities is less constrained than the inferred temperatures, but still, through a large part of the domain, more constrained than the velocities in the simulation. 

\subsection{Estimating chromospheric heating}
Since our simulation contains a physically consistent atmospheric solution following rMHD, our method makes it straightforward to study the terms in the energy equation. In this section we study the dissipative chromospheric heating, and test whether it is viable to estimate chromospheric heating from the \theline\ line using our database.

\begin{figure}
    \resizebox{\hsize}{!}{\includegraphics{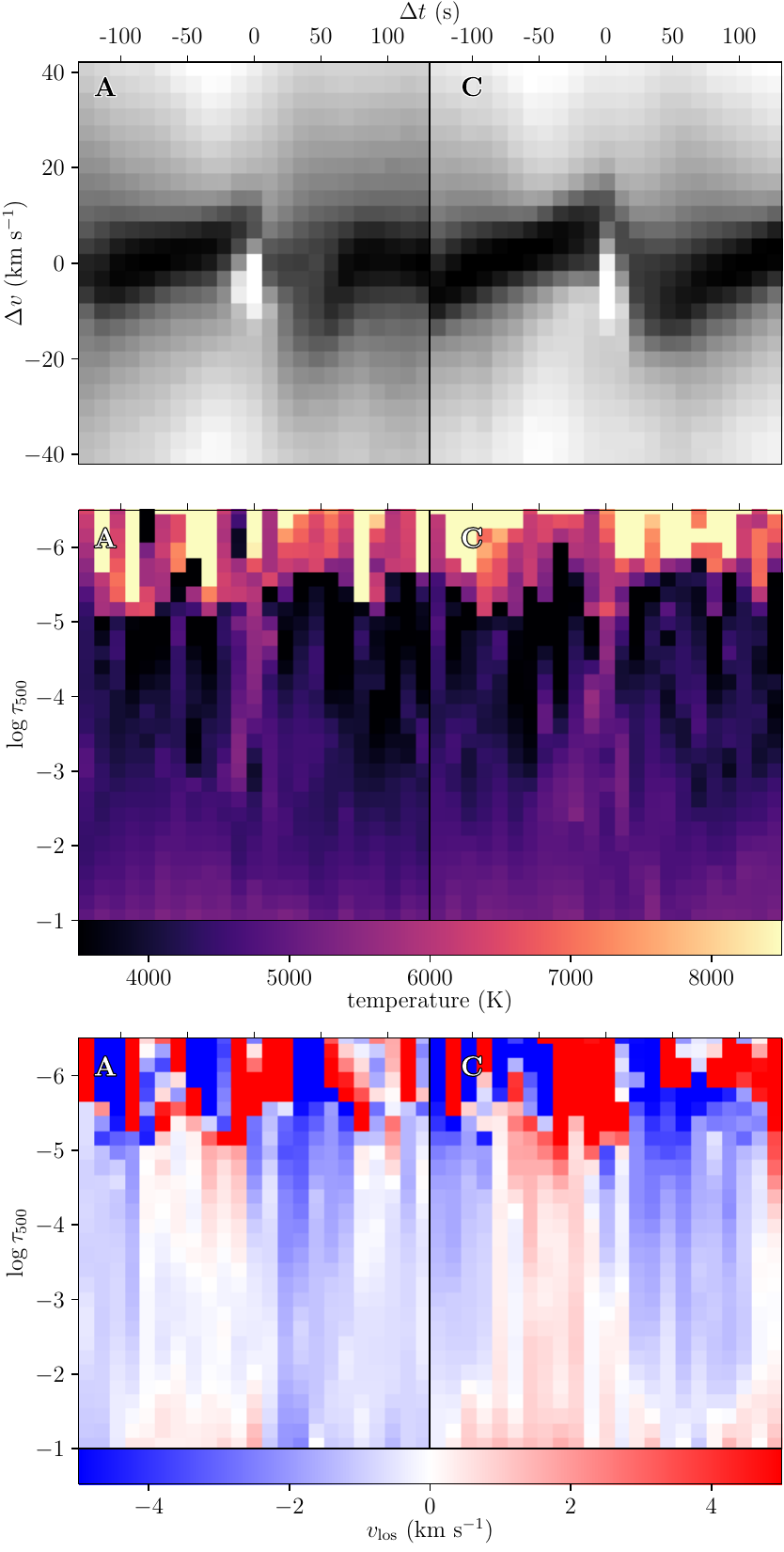}}
    \caption{Inversion of events A and C marked in Fig.~\ref{fig:heating}. \textit{Top}: Spectrograms from the observation. \textit{Middle}: Time--distance evolution of the temperature. \textit{Bottom}: Time--distance evolution of the line-of-sight velocity. $\Delta t = 0$ corresponds to the same time as the frame at 2014-06-24T09:15. The sign of velocities follows the same convention as in Fig.~\ref{fig:inversion_tau500}.}
    \label{fig:tseries}
\end{figure}

We calculated the dissipative heating $Q_\mathrm{diss.} = Q_\nu + Q_\eta$ for the same frame shown in Fig.~\ref{fig:observation_overview} and integrated this term over the column mass range from \qtyrange{1e-1}{1e-3}{kg.m^{-2}} to obtain the total lower chromospheric heating $Q_{\rm lower}$. The integrated lower chromospheric heating is shown in Fig.~\ref{fig:heating}. $Q_{\rm lower}$ is strong in the network, with typical values around \qty{10}{kW.m^{-2}}. In comparison, the field-of-view-averaged heating through the lower chromosphere is \qty{3.3}{kW.m^{-2}}. However, the highest estimated chromospheric heating occurs in small isolated regions that appear bright in the line core and show no noticeable features in the far wings (pseudo-continuum) of the line. We have labelled some of these high heating locations in the figure; all have spectral signatures corresponding to bright blue grains. Locations A and C reach integrated chromospheric heating rates just above \qty{50}{kW.m^{-2}} and location B has a value of \qty{25}{kW.m^{-2}}. Locations A and C are the two areas with the strongest estimated heating in the inverted frame.

The uncertainty of our heating estimates is shown in the histogram on top of Fig.~\ref{fig:heating}. The uncertainty is on average 40\%, and the distribution of uncertainties is close to normal. The uncertainties are quite high, and the distribution of the uncertainties is only a little more constrained than the spread of heating values in the simulation. Therefore, the inferred heating values have significant uncertainties, even though the map of heating values appears sensible, with the highest heating values located in bright blue grains and the network.

\subsection{Time series analysis}
We investigated the temporal evolution of the bright grains labelled A and C in Fig.~\ref{fig:heating}. The spectrograms and inverted atmospheric quantities from these events are shown in Fig.~\ref{fig:tseries}. The spectrograms show clear blue-grain signatures at $\Delta t = 0$: the line cores oscillate in a sawtooth pattern leading up to the event and emission in the inner blue wing occurs at $\Delta t = 0$. The observations are well reconstructed with BiSPEC, with the only visible difference being a smoother signal in the reconstructed spectrograms. 

\begin{table}
\caption{Nodes used for the STiC inversions.}
\label{tab:stic}
\centering
\begin{tabular}{c c c c}
\hline\hline
Parameter & cycle 1 & cycle 2 & cycle 3 \\
\hline
    temperature           & 4 & 9 & 9 \\
    $v_\mathrm{los}$   & 1 & 4 & 8 \\ 
    $v_\mathrm{turb}$  & 1 & 4 & 8 \\
\hline
\end{tabular}
\end{table}

The time-distance diagrams in Fig.~\ref{fig:tseries} show the atmospheric responses to the bright blue grains. The temperature shows an increase through the entire chromosphere in both events. Both events are approximately \qty{1500}{K} hotter at $\ltau = -5$ than the typical temperature at this depth. Event A is a few hundred kelvins hotter than event C lower in the atmosphere. In the line-of-sight velocity, we see evidence of upward-propagating waves in the chromosphere. This is most clearly seen around $\ltau = -5$ at $\Delta t = 0$ where, for both events, upflowing material meets downflowing material from above and there is a large gradient in the velocity. In the lower parts of the chromosphere the velocity amplitude is smaller. It is worth noting that the uncertainty in the line-of-sight velocity above $\ltau = -5$ is high, so the values at these depths are less reliable than those lower in the atmosphere.

\subsection{Comparison with STiC}
We inverted the small $11\farcs4 \times 11\farcs4$ patch that is indicated in the top row of Fig.~\ref{fig:observation_overview} with the STiC inversion code. The patch was placed so that it covered only a quiet-Sun region. STiC was run for three cycles with the parameters shown in Table~\ref{tab:stic}.

\begin{figure}
    \resizebox{\hsize}{!}{\includegraphics{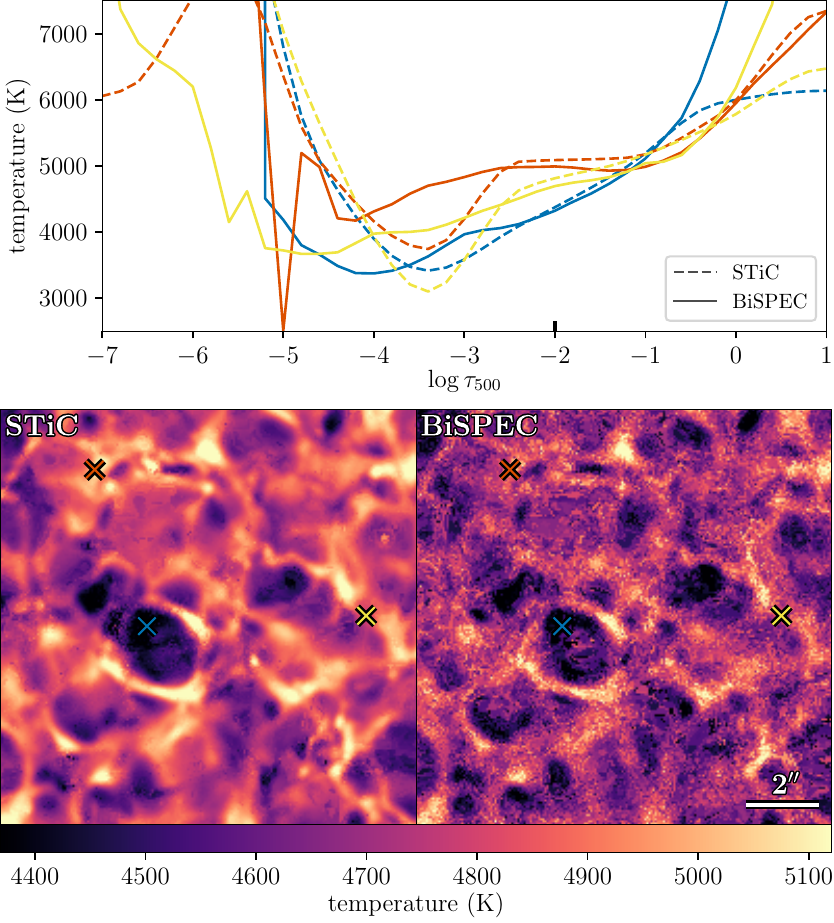}}
    \caption{Comparison between STiC and BiSPEC temperature inversions. The inversions are performed on the smaller patch shown in Fig.~\ref{fig:observation_overview}. \textit{Top row}: Inverted temperatures from a few selected pixels, which are marked with crosses in the bottom row. \textit{Bottom row}: Inverted temperature obtained with STiC (left) and BiSPEC (right) at $\ltau = -2$. The depth at which we compare is marked on the $x$-axis of the top plot.}
    \label{fig:compare_temp}
\end{figure}

Results from the STiC inversion are plotted and compared with the BiSPEC inversion in Figs.~\ref{fig:compare_temp} and \ref{fig:compare_vlos}. Figure~\ref{fig:compare_temp} compares the temperature inversions. The STiC and BiSPEC temperature stratifications in the top panel show similar trends in the lower atmosphere for the few pixels plotted. The temperature maps at $\ltau = -2$ in the lower panel show that STiC generally estimates higher values than BiSPEC. Most of the locations are a few hundred kelvins hotter in the STiC inversion, but in return the hot locations in the BiSPEC inversion are hotter compared with the STiC inversion. Therefore, the average temperature over the entire patch derived via the two methods is almost the same.

Figure~\ref{fig:compare_vlos} compares the inverted line-of-sight velocities between STiC and BiSPEC similarly to Fig.~\ref{fig:compare_temp}. There is, unlike the temperature, little agreement in the line-of-sight velocity derived using the two inversion methods, which is highlighted by the stratifications of the selected pixels that are plotted in the top row. However, in the map of velocities in the bottom row, we can see some similar trends between STiC and BiSPEC. Many locations agree on the sign of the velocity, but BiSPEC estimates lower amplitudes in the velocity than STiC. The smaller amplitudes we see with BiSPEC partly come from the smoothing effect that the $k$-NN inversion has on this quantity.

Since we used only \theline\ for inversions, temperatures and line-of-sight velocities outside the sensitivity of this line are poorly constrained with STiC. On average, the STiC inferred temperatures and line-of-sight velocities show unexpected behaviour outside the range $\ltau = -1$ to $\ltau = -3$. There are usually large downflows below and large upflows above this range, and individual columns do not exhibit coherent spatial structures in either velocity or temperature outside the same range.

\section{Discussion}
We derived atmospheric parameters from solar \theline\ observations by using a synthetic spectra database that was built from a 3D rMHD simulation. This work is a first approach to inverting chromospheric spectra with a Bifrost synthetic database, and therefore we chose to keep the inversion method as simple and explainable as possible while still obtaining reasonable results from our inversions. To this end, we used $k$-NN inversion with $k=15$ neighbours, which yielded inversions more consistent than with nearest-neighbour interpolation. The $k$-NN used a Euclidean distance metric that weighted all line positions equally, and we then computed the inversion as the medoid of the 15 closest neighbours. The number of neighbours was chosen by visually inspecting the results at different values of $k$, and our method was justified by validating our inversion on \theline\ spectra generated from a different simulation. 
The validation showed an excellent estimation of the gas temperature, with average errors around a few percent between $\ltau = -3$ and $\ltau = -0.5$. The inverted line-of-sight velocity showed qualitative agreement with the ground truth, but the velocity-amplitude was generally much weaker than in the simulation. 

\begin{figure}
    \resizebox{\hsize}{!}{\includegraphics{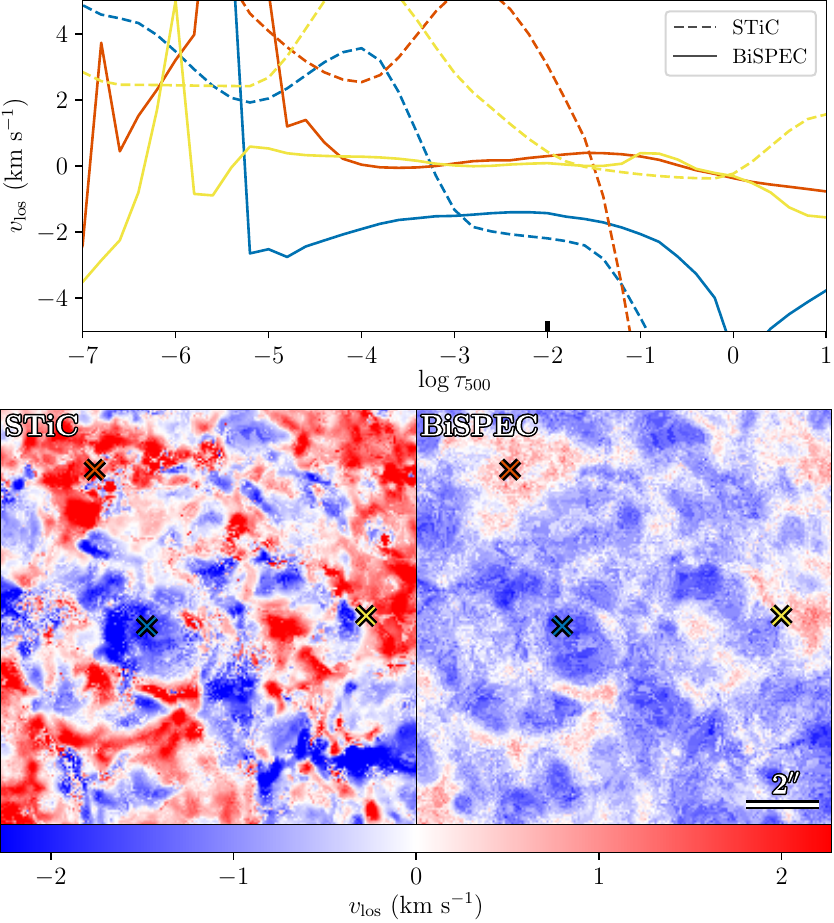}}
    \caption{Same as Fig.~\ref{fig:compare_temp} but for line-of-sight velocity. Positive values (red) correspond to downflows, and negative values (blue) to upflows.}
    \label{fig:compare_vlos}
\end{figure}

An intrinsic limitation of pixel-by-pixel inversions is the degeneracy of the formation of radiation. We saw that columns that produced similar \theline\ profiles also had similar stratifications in optical depth (see the example in Fig.~\ref{fig:knn_regression}) but not on a physical height scale. While optical-depth (or column-mass) scales are the standard measure for height in pixel-by-pixel inversions, recent methods using machine learning have been able to resolve the degeneracy in the formation of spectra and recover the geometric height in inversions, for example by using convolutional neural networks \citep{Asensio-Ramos:2019aa} or physics-informed neural networks \citep{Yang:2025aa}. Breaking the formation degeneracy and recovering quantities on the original geometric-height scale of the simulation is a natural next step to explore for a simulation-informed database inversion such as BiSPEC.

BiSPEC provides a physically consistent solution for a large range of $\tau_{500}$ optical depths. In the line-of-sight velocity, we saw consistent but slightly underestimated results from the photosphere to above the temperature minimum ($-5 \lesssim \ltau \lesssim 0$), and in temperature we recovered sensible estimates between $-4 \lesssim \ltau \lesssim -1$. Both of these ranges, in particular the larger optical depths, were beyond the sensitivity of the \theline\ line positions we used, which means they cannot be constrained by the data. For locations outside the line sensitivity, BiSPEC is in effect providing an extrapolated solution according to the first-principle physics of the simulations. STiC, on the other hand, does not follow such constraints and therefore its estimates outside the line sensitivity region are often poorly constrained, especially at higher optical depths. However, our method is still bound by the line sensitivity, and uncertainties in our derived quantities are constrained by the information coming from a single line. Including more spectral lines to constrain the database inversions and exploring different weightings of spectral ranges will improve the range of inversions (e.g. as done by \citealt{Sainz-Dalda:2026aa} to improve the IRIS$^{2+}$\;sensitivity to the broader range $ -7.6 < \ltau < 0$), both from our database method and from STiC. 

We calculated the spread in our inversions from the $k$-NNs for each pixel in the inverted frame and used this as a gauge of uncertainty and of the degree of degeneracy in our inversion. Our estimates are limited by the sample space of the simulation and the number of neighbours that can accurately reproduce the observations. We found that both the temperature and the line-of-sight velocity were best constrained for $-3 < \ltau < -1$. RADYNVERSION \citep{Osborne:2019aa} also uses \theline\ for inversions and estimates low uncertainties in the lower chromosphere. However, in this range our values for temperature were also concentrated around their median values, as shown in Fig.~\ref{fig:inversion_deviation}, and is a point that is also made by \citet{Schmit:2021aa}. By comparing the uncertainties in our inversions with the spread of the same quantity in our simulation, we found that for some optical-depth ranges, inverted temperatures and line-of-sight velocities had significantly lower uncertainties than the spread of values in the simulation. This means that our inferred values are constrained by the data at these optical depths. For the temperature, the inversions are meaningful between $-3 < \ltau < -1$, while the line-of-sight velocity is informed from the data all the way from the photosphere at $\ltau = 0$ until above the average depth of the temperature minimum at around $\ltau \approx -4$. Also in this case, including more spectral lines can be beneficial in order to reduce uncertainties and resolve degeneracies in the inversion \citep{Schmit:2021aa}. 

We saw some noise in our inversion, which was especially apparent in the zoomed-in field of view in Figs.~\ref{fig:compare_temp} and \ref{fig:compare_vlos}. Using several spectra to fit the atmospheric quantities created a solution that is smoother than a simple nearest-neighbour interpolation (refer to Appendix \ref{app:k-nearest}), but pixel-by-pixel inversion is generally noisy. There exist other strategies to smooth inversions, which we did not employ here. A key strategy would be to include spatial information in the inversion. Spatial regularisation is a technique that has been proven advantageous in spectropolarimetric inversions \citep{Morosin:2020aa,Diaz-Baso:2025aa}, and would be a natural step to  explore in the future. Another capability of the simulation that we did not exploit is the spatial structure in the synthetic spectra. With the simulation, we have the capability to encode spatial information into the inversion, for example by using spatial convolution. Encoding spatial information into the inversion could break the degeneracy of the formation of the spectra and make it possible to obtain the thermodynamical variables at a geometric height scale. Spatially informed inversions \citep[e.g. inversions using convolutional neural networks,][]{Milic:2020aa} also show less noise than pixel-by-pixel inversions. Therefore, while the simple first-approach we present here is already powerful, we think there is a substantial potential for improvement by incorporating spatially informed inversion strategies.

One of the main reasons to use database inversions is the speed-up compared to traditional inversion codes. With a single workstation, our method inverts a single pixel in about \qty{40}{ms} per core, which was almost $10^4$ times faster than the STiC inversion. Other works have seen tremendous speed-ups with database inversions based on a nearest neighbour approach \citep[e.g.][]{Beck:2015aa,Sainz-Dalda:2019aa,Beck:2019aa,Sainz-Dalda:2026aa}. \citet{Raja-Bayanna:2026aa} show that database inversions show only minor differences compared to pixel-by-pixel inversions in active region observations while being quick enough to invert large datasets. \citet{Asensio-Ramos:2025aa} also used a database approach to inversions, but they utilised a machine-learning method that provides the probability for each model atmosphere and estimates uncertainties in the inversions. However, since those works are based on modified model atmospheres, their models do not have the same level of variation or physically consistent sharp gradients that a realistic rMHD simulation can offer. \citet{Riethmuller:2017aa} performed database inversions to evolve an rMHD simulation and a physically consistent atmosphere. Unlike in our work, they used only the single best matching synthetic spectrum for the inversion, and they were limited to photospheric inputs since their simulation did not include a chromosphere. 

An advantage of using rMHD simulations for inversions is the fact that we sample line profiles that incorporate a realistic description of the underlying physics. The simulation features are consistent with what we see in the observations: granule spectra from granule regions, and the same for intergranular lanes, bright grains, and so on. In comparison with standard inversion routines, BiSPEC does not rely on parametrised broadening terms (microturbulence), and uses dynamical atmospheres (i.e. without assuming hydrostatic equilibrium). Missing from the current simulation is a more realistic model of network regions, and we are therefore careful not to make strong conclusions regarding these areas. This is a weakness of the database and not of the method, and can be easily mitigated by including spectra from an array of different simulations \citep[e.g. the enhanced network simulation in][which has network features]{Carlsson:2016aa}.

Finally, as a proof of concept, we estimated the chromospheric heating and temporal evolution associated with two bright blue grains. We reconstructed the line-of-sight velocity and saw evidence of upward-propagating shocks, but with lower velocity amplitudes than those reported in similar Bifrost simulations \citep{Udnaes:2025aa,Udnaes:2026aa} and in inversions performed with STiC \citep{Mathur:2022aa}. Although the inverted spectra match the observed ones, the corresponding atmospheric profiles do not necessarily reproduce the same spectra when used in forward synthesis, since an average spectrum does not correspond to an average atmosphere \citep{Uitenbroek:2011aa}. One factor leading to the smaller velocity amplitudes is the degeneracy in the formation of \theline, but the spread that we infer in the line-of-sight velocity is too small to account for the large difference in amplitudes compared to the simulation. We therefore assume that the spectral degradation applied to the synthetic spectra also contributes to this discrepancy. The lower chromospheric heating in the bright blue grains was around an order of magnitude higher than the average, with a temperature increase of around \qty{1500}{K}. This temperature increase is comparable to the average temperature excursions from bright blue grains in our simulation \citep[\qty{2500}{K} for the strongest blue grains in the same simulation,][]{Udnaes:2026aa}, and also to values previously inferred from SST observations \citep[\qtyrange{1000}{2000}{K} in][]{de-la-Cruz-Rodriguez:2013aa,Joshi:2018aa,Mathur:2022aa}.

\section{Conclusions}
We have presented a method for inverting \theline\ observations using a synthetic-spectra database. While this idea builds on previous work, this is the first time such a method has been applied to chromospheric spectra from a 3D rMHD simulation. Our method also differs from previous work in that it uses a $k$-NN inversion algorithm instead of nearest-neighbour interpolation. By inverting an observation as the medoid of the 15 nearest neighbours in our database, we retrieved a solution that is spatially more consistent than with nearest-neighbour interpolation, and we used these nearest neighbours to estimate the uncertainty in the inversion.

The database was sampled from 150 million \theline\ spectra synthesised in non-LTE from a Bifrost simulation. Atmospheric stratifications were inferred from the simulation columns of the 15 nearest-neighbour synthetic spectra. We validated our inversion with spectra from another unseen simulation and found excellent agreement in the gas temperature. Using SST observations,  we inverted the temperature and line-of-sight velocity onto a common $\tau_{500}$ depth scale. We found that the temperature was well constrained with our inversion method, while the line-of-sight velocity generally had a higher level of degeneracy. Temperature inversions were reliable in the range $-4 \lesssim \ltau \lesssim -1$, while the line-of-sight velocity was recovered over a much larger optical-depth range from the photosphere to the temperature minimum ($-5 \lesssim \ltau \lesssim 0$). Comparing the uncertainties of our inversions with these quantities' distributions in our simulation, we found that the temperature inversions contained more information on the inverted distributions in the range $-3 \lesssim \ltau \lesssim -1$, and the line-of-sight velocity was similarly more constrained than the model in the range $-4.5 \lesssim \ltau \lesssim 0$.

Compared to traditional inversion methods, our database inversion is much faster once the spectra have been synthesised, requiring only \qty{40}{ms} per pixel without trimming the database in any way. Compared to STiC, our inversions also provide a physically consistent extrapolated solution outside the sensitivity range of the spectral line, since all inverted pixels are a column from a magnetohydrodynamic simulation. The temperatures we inverted agree well with those determined using the traditional inversion method of STiC, while our line-of-sight velocity estimates have lower amplitudes. 

Finally, we tested the method by inspecting results for two bright blue grains. We recovered reasonable results, with a temperature rise consistent with previous studies and signatures of vertically propagating shock waves. The inverted velocities were lower than expected from previous analyses of simulations and comparisons with other inversions, highlighting current limitations of our method. 

As a first approach to inverting chromospheric spectra with 3D rMHD simulations, our method is conceptually simple and computationally efficient. It recovers physically consistent results, free from simplifying assumptions such as hydrostatic equilibrium or the fudge factor of microturbulence, and allows for naturally occurring sharp gradients in quantities without the trade-offs of a node-based inversion (e.g. artificially smooth features or overfitting). It also comes with error estimation built in, which can provide observers with a crucial gauge of how reliable a given inversion is. Our method inherits the limitations of the underlying simulation, but this will improve as simulations get better. 

We propose future improvements that we believe will mitigate many of the challenges we have identified:
\begin{itemize}
    \item Including additional spectral lines with different formation heights and sensitivities to constrain temperatures and velocities over a larger range of optical depths.
    \item Exploiting the spatial structure of the simulation in the inversion, for example with spatial regularisation or machine-learning methods using convolution (e.g. to recover the inversions on a geometric-height scale).
    \item Expanding the database with spectra from a diverse set of simulations that represent different solar features, for example the network regions seen in our observations.
\end{itemize}
As rMHD simulations become increasingly realistic, advances in database inversions have a large potential to strengthen chromospheric diagnostics.

\begin{acknowledgements}
We thank Carlos J. Díaz Baso for useful discussions related to the method.
This research has been supported by the Research Council of Norway through its Centres of Excellence scheme, project number 262622, and by computational resources provided by Sigma2, the National Infrastructure for High Performance Computing and Data Storage in Norway. TMDP's work benefited from discussions and support by the International Space Science Institute project (ISSI-BJ ID 24-604) on ``Small-scale eruptions in the Sun''.
The Swedish 1-m Solar Telescope (SST) is operated on the island of La Palma by the Institute for Solar Physics of Stockholm University in the Spanish Observatorio del Roque de los Muchachos of the Instituto de Astrof{\'\i}sica de Canarias. The SST is co-funded by the Swedish Research Council as a national research infrastructure (registration number 4.3-2021-00169).
\end{acknowledgements}

\bibliographystyle{aa}
\bibliography{references}

\begin{appendix}

\section{$k$-nearest neighbours inversion} \label{app:k-nn}
In this appendix we provide a detailed description of the $k$-NN method employed in our inversion. The $k$-NNs for a \theline\ observation (labelled with the index $j$) are found by:

\begin{enumerate}
    \item Normalising the observed profile $I^{\rm o}_j$ and all synthetic profiles $I^{\rm s}_l$ by their respective spatially and spectrally averaged intensities,
    \begin{equation}
        \tilde I^{\rm o}_j \;=\; 
        \frac{I^{\rm o}_j}{\mathrm{mean}(I^{\rm o})}
        \quad\text{and}\quad
        \tilde I^{\rm s}_l \;=\;
        \frac{I^{\rm s}_l}{\mathrm{mean}(I^{\rm s})}\,.
    \end{equation}
    Here, $\mathrm{mean}(I^{\rm o})$ and $\mathrm{mean}(I^{\rm s})$ denote the mean intensity over all spatial positions and wavelengths in the observed and synthetic datasets, respectively.

    \item Computing the distance between the normalised observation and each normalised synthetic spectrum in the database. For each synthetic profile with index $l$, the squared distance $d_{j,l}^2$ is defined via the $L_2$ norm as
    \begin{equation}
        d_{j,l}^2 \;=\;
        \sum_{i=1}^{n_\lambda}
        \left(
            \tilde I^{\rm o}_{j,i} - \tilde I^{\rm s}_{l,i}
        \right)^2,
    \end{equation}
    where $i$ indexes wavelength and $n_\lambda$ is the number of wavelength positions.

    \item Constructing the set $\mathcal{N}_j$ of the $k$-NNs of the observation $j$ by selecting the synthetic spectra with the $k$ smallest distances $d_{j,l}$.

    \item Computing the inversion as the medoid over these $k$-NNs. The medoid is computed from the concatenated stratifications of both the temperature and line-of-sight velocity of each simulation column in the set of nearest neighbours. In the concatenation, both quantities are $Z$-normalised, and the medoid is computed with the L1 norm.
\end{enumerate}

The distance calculations are supported by a k-D tree; we used the {\tt KDTree} module \citep{Maneewongvatana:1999} from the {\tt scipy} Python package \citep{2020SciPy-NMeth}. The indices of the nearest neighbours are easily found with just a few lines of code:
\begin{verbatim}
    tree = KDTree(database)
    distances, indices = tree.query(
        observation, 
        k=15
    )
\end{verbatim}
where the {\tt database} is the synthetic spectra matrix with shape $(N, n_\lambda)$.

\begin{figure}
    \resizebox{\hsize}{!}{\includegraphics{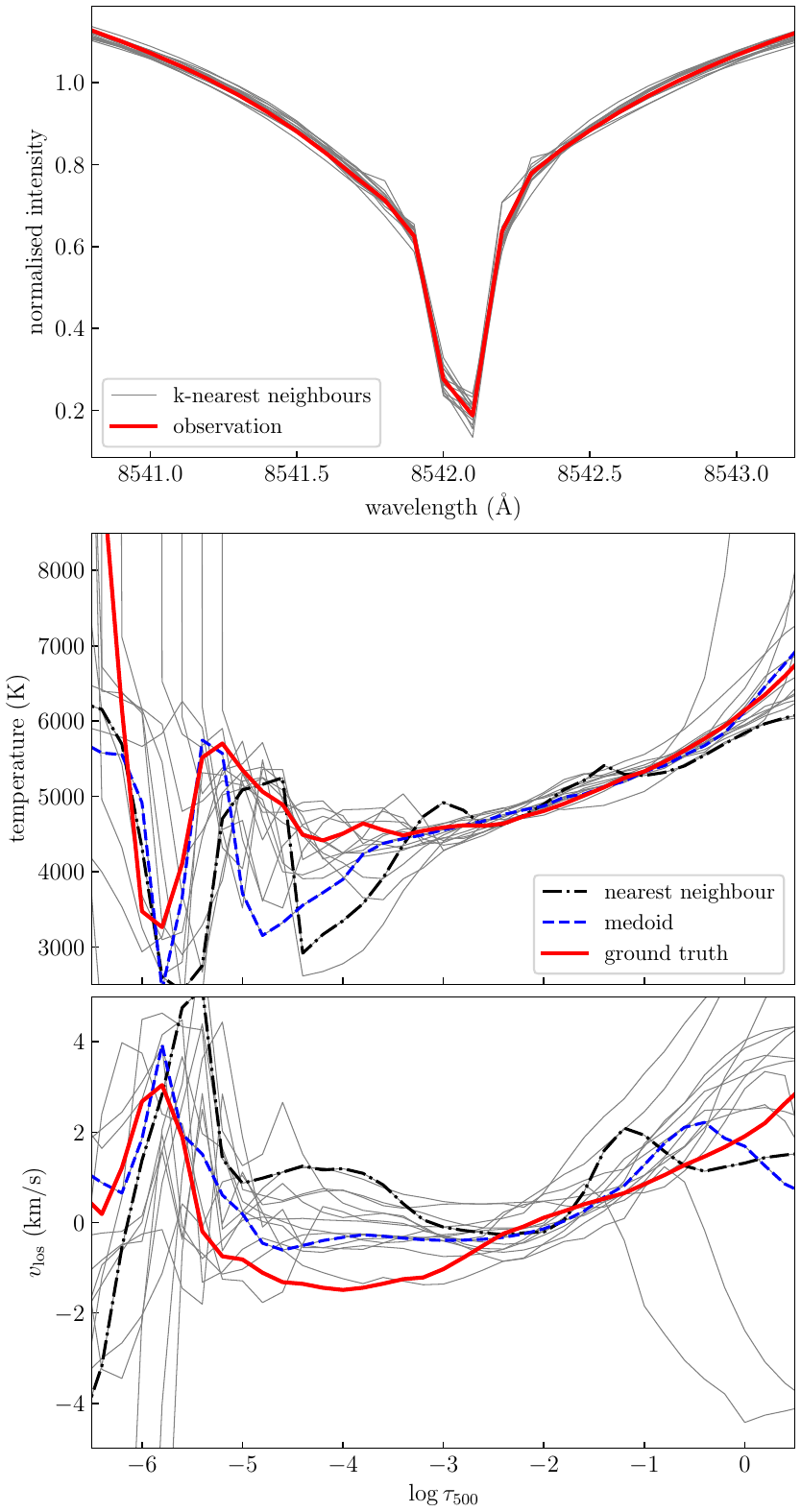}}
    \caption{Synthetic spectra and atmospheric stratifications for the $k$-NNs to an observation from the enhanced network simulation. \textit{Top}: Synthetic spectra. The thick red line marks the synthetic observation, and the thin grey lines are the $k$-NN spectra from the database. \textit{Middle} and \textit{bottom}: Temperature and line-of-sight stratifications of the $k$-NN. The stratifications are compared to the ground truth taken from the enhanced network simulation (red line). Stratifications from the nearest neighbour and medoid are highlighted.}
    \label{fig:knn_regression}
\end{figure}

The inversion routine for a single observation is exemplified in Fig.~\ref{fig:knn_regression}. The top panel shows an observation from the enhanced network simulation that is reproduced by our database. In the lower panels we show the simulation columns associated with the observation and the $k$-NN, and highlight both the nearest neighbour and the medoid. The simulation temperatures have similar stratifications in the lower atmosphere, which are represented well by the medoid. The velocities have a larger spread, which is expected, and deviate from the medoid throughout the domain.

\clearpage

\section{Comparison with nearest-neighbour interpolation} \label{app:k-nearest}

\begin{figure}
    \resizebox{\hsize}{!}{\includegraphics{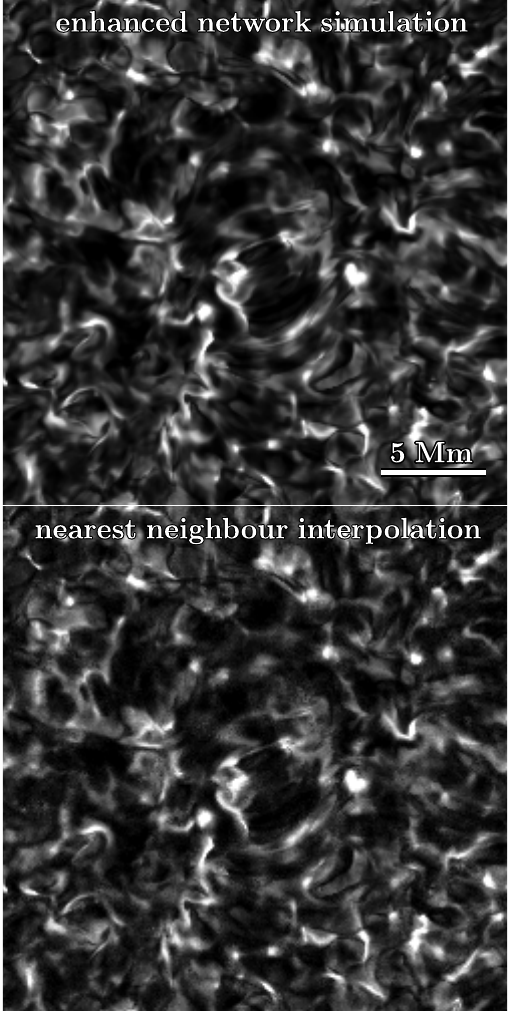}}
    \caption{Line core of \theline\ from the enhanced network simulation (\textit{top}) compared to the reconstructed intensity with nearest-neighbour interpolation from the synthetic spectra database (\textit{bottom}).}
    \label{fig:compare_k}
\end{figure}

In this section we show the difference between $k$-NN regression and nearest-neighbour interpolation. Nearest-neighbour interpolation is analogous to $k$-NN regression with $k=1$, where the inversion is simply taken from the closest synthetic spectrum in our database. Figure~\ref{fig:compare_k} shows the \theline\ line core from the enhanced network simulation with the reconstructed line core intensity from the spectral database using the nearest-neighbour interpolation method. The reconstructed intensities show a good qualitative agreement with the `observation', but the reconstructed image is grainy in some locations.

\begin{figure*}
    \includegraphics[width=17cm]{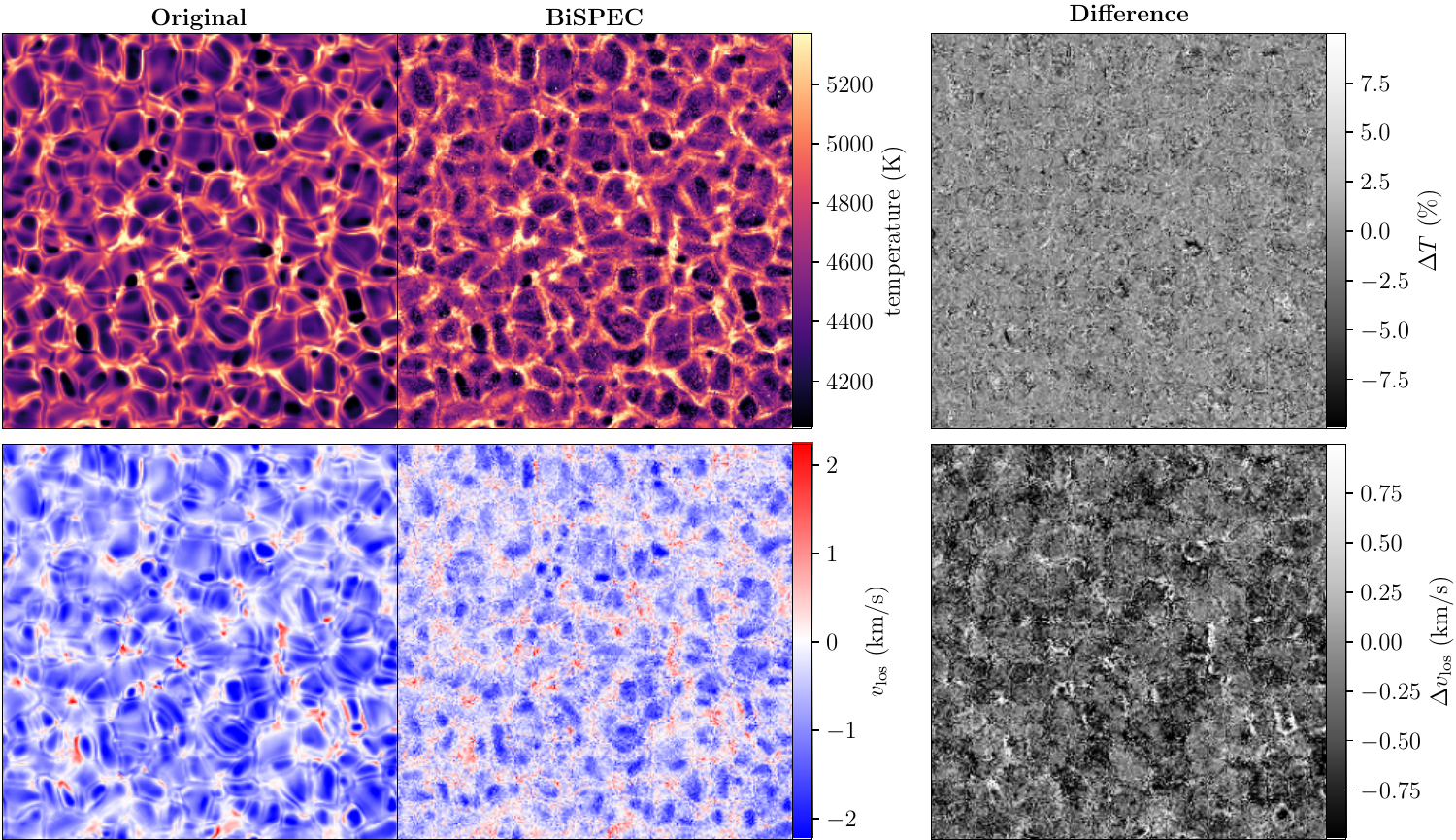}
    \caption{Same as Fig.~\ref{fig:sim_comp} but the inversions are calculated with nearest-neighbour interpolation. The inversions are plotted at $\ltau = -1.8$.}
    \label{fig:inversion_nearest_neighbours}
\end{figure*}

Figure~\ref{fig:inversion_nearest_neighbours} shows the inverted atmosphere of the enhanced network simulation from the nearest-neighbour interpolation, where the spatial maps of atmospheric variables are noisier than the $k$-NN regression shown in Fig.~\ref{fig:sim_comp}. The line-of-sight velocities generally have larger amplitudes, since this method does not have any smoothing. Still, the spatial structure here is less apparent than with $k$-NN regression. 

\end{appendix}

\end{document}